\documentclass{article_saj}
\usepackage{graphicx,saj,multicol,subeqnarray}
\usepackage{natbib}
\usepackage{float}
\usepackage{xcolor}
\usepackage{widetext}
\usepackage{url}
\usepackage{bm}
\usepackage{tikz} 
\usepackage{pifont} 
\usepackage{amsfonts}
\usepackage{amssymb}
\usepackage{amsmath,upgreek}
\usepackage{titlesec}
\titlelabel{\thetitle.\quad}
\definecolor{xlinkcolor}{cmyk}{1,0.6,0,0}
\usepackage[bookmarks=false,         
     pdfnewwindow=true,      
     colorlinks=true,    
     linkcolor=xlinkcolor,     
     citecolor=xlinkcolor,     
     filecolor=xlinkcolor,  
     urlcolor=xlinkcolor,      
final=true
]{hyperref}

\renewcommand{\thefootnote}{\fnsymbol{footnote}}

\begin{document}
\parindent=.5cm
\baselineskip=3.8truemm
\columnsep=.5truecm
\newenvironment{lefteqnarray}{\arraycolsep=0pt\begin{eqnarray}}
{\end{eqnarray}\protect\aftergroup\ignorespaces}
\newenvironment{lefteqnarray*}{\arraycolsep=0pt\begin{eqnarray*}}
{\end{eqnarray*}\protect\aftergroup\ignorespaces}
\newenvironment{leftsubeqnarray}{\arraycolsep=0pt\begin{subeqnarray}}
{\end{subeqnarray}\protect\aftergroup\ignorespaces}

\begin{strip}

{\ }

\vskip-1cm

Initial Version: October 16, 2024 \newline
Current Version: November 25, 2024

{\ }
\title{CONCENTRATED SUPERELLIPTICAL MARKET MAKER}
\vskip3mm
\authors{Vasily Tolstikov}
\vskip3mm
\vskip3mm
\Email{0xeb567e5ac2feb7bbfc777e0bd18a8a34eb922e86@ethereum.mailchain.com}
\vskip3mm

\summary{
An automated market maker where the price can cross the zero bound into the negative price domain with applications in electricity, energy, and derivatives markets is presented. A unique feature involves the ability to swap both negatively and positively priced assets between one another, which unlike traditional markets requires a numeraire in the form of a currency. Model extensions to skew and concentrate liquidity are shown. The liquidity fingerprint, payoff, and invariant are compared to the Black-Scholes covered call and the Logarithmic Market Scoring Rule invariants.
}
\keywords{Automated market maker, Black-Scholes, Constant product market maker, Concentrated liquidity, Heavy tails, Lam\'e curve, Legendre Transform, Logarithmic Market Scoring Rule, Liquidity fingerprint, Negative price, Negative liquidity.}

\end{strip}

\tenrm

\section{INTRODUCTION}

\indent

\footnotetext[0]{Acknowledgements: special thanks to Dan Robinson for invariant and liquidity fingerprint discussions.}
Negative prices are an unusual phenomenon in traditional financial markets. As the set of Real World Assets (RWA) is tokenized and becomes tradeable on the blockchain, the probability of encountering a negative price approaches one hundred percent. Currently, we do not have the ability
to provide liquidity or trade an asset on the blockchain with a negative price on an AMM. This paper:

\par\hang\textindent{(1)} Outlines the phenomenon of negative prices in financial literature. 

\par\hang\textindent{(2)} Reviews the limitations of current AMMs and how to enable negative prices.

\par\hang\textindent{(3)} Compares the payoffs and proposes unique applications of a negative price AMM.

\renewcommand{\thefootnote}{\arabic{footnote}}

\parindent=.5cm
\section{DISCUSSION}

\indent

The phenomenon of negative prices has been observed empirically, especially in commodity markets. Examples being
oil in 2020 [\hyperlink{ref-1}{1}], onions in 1955 [\hyperlink{ref-2}{2}],  electricity in the US in 2024 [Figure1A] [\hyperlink{ref-3}{3}], fat-finger trading mistakes in Finland in 2023 pushing prices negative [\hyperlink{ref-4}{4}], and synthetic
financial products consisting of derivatives of highly liquid US treasury bonds [\hyperlink{ref-5}{5}].

In the world of traditional finance, price manipulation has gone so far as to encourage Congress to ban commodities that have gone negative in price as was the case in 1958 with the passage of the Onion Futures Act, effectively banning the trading of onions to this day due to the bankruptcies caused by onion price manipulation [\hyperlink{ref-6}{6}]. Unlike traditional finance, the permissionless nature of decentralized ledgers allows for the exchange of any token, whether banned or not, but the feature of a negative price is not possible with current AMM invariants. 
\subsection{AMM Invariant}

Uniswap v2 introduced a continuous uniform liquidity distribution [\hyperlink{ref-7}{7}] allowing for the trading of a token in the price range of zero to infinity. Yet prices are not infinite nor bound by the zero barrier, to address part of this problem Uniswap v3 was introduced to allow for the concentration of liquidity based on an LP's preferred viewpoint on the range of price movement between zero and infinity [\hyperlink{ref-8}{8}]. The constrained nature of a discrete uniform liquidity distribution of a Uniswap v3 position creates a problem of capital inefficiency in the tails where an LP can wind up out of range while earning no fees [Figure2]. To deal with being out of range, multiple Uniswap v3 positions can be customized along curves or, alternatively, AMM invariants can be made that match particular price behavior such as Geometric Brownian Motion in Black Scholes with Replicating Market Makers (RMM) [\hyperlink{ref-9}{9}].This approach is extended by making the empirical observation that real world financial assets, as they continue to migrate onto the blockchain, can exhibit heavy tails [Figure1B][\hyperlink{ref-10}{10}][\hyperlink{ref-11}{11}] as well as enter the negative price domain [\hyperlink{ref-12}{12}].

It should be noted that Uniswap's Constant Product Market Maker (CPMM) invariant does have a negative liquidity domain (see Appendix A), but accessing it is limited due to the nature of the hyperbolic equation $xy=k$.  We adjust the Uniswap v3 AMM invariant equation below by allowing the curve to fold in on itself to access negative prices and the liquidity in the negative domain.
\begin{equation}
\left(x-a\right)^{2}+\left(y-b\right)^{2}=k^{2}
\label{eq:1}
\end{equation}
where $a$ and $b$ are offsets equal to $k$ to pin liquidity. The liquidity fingerprint of this Concentrated Circular Market Maker (CCMM) is derived in Appendix B and can be extended along a Lam\'e curve [\hyperlink{ref-13}{13}] to create a Concentrated Super-Elliptical Market Maker (CSEMM) to further widen/narrow and skew the liquidity allocation with the invariant becoming:
\begin{equation}
\left|\frac{x}{\alpha}-1\right|^{u\left(\alpha\right)}+\left|\frac{y}{\beta}-1\right|^{u\left(\beta\right)}=1
\label{eq:2}
\end{equation}
where $u(x)=\frac{ln(2)}{ln(\frac{x}{x-1})}$. Parameters $\alpha$ and $\beta$ allow for the extension of the liquidity in the left and right tail. By setting $\alpha=\beta$ and decreasing their values towards $2$ one can approach the liquidity concentration of $x+y=k$, by widening $\alpha$ and $\beta$ towards infinity one increases the tails of the distribution until it converges to the Logarithmic Market Scoring Rule (LMSR) invariant. At $\alpha=\beta=\frac{1}{\sqrt{\frac{\left(\sqrt{2}-1\right)^{2}}{2}}}$ we retrieve the CCMM from equation 1.
\newline
\indent
The swap functions are given in Appendix C, compared to other invariants in Figure 3 and the RMM with the LMSR in desmos: \url{https://www.desmos.com/calculator/mbohmvmytg}. Its unique feature allows one to trade two assets where either price can go negative between one another. While appearing bizarre, it is normal to be able to exchange a barrel of oil costing negative four dollars for two natural
gas costing a negative two dollars each. This is at the present moment not doable in centralized exchanges, which
in the middle of COVID-19 had difficulty accounting for negative oil prices by having to switch to Arithmetic Brownian
Motion from Geometric Brownian Motion via the Bachelier Model [\hyperlink{ref-14}{14}]. Furthermore, for an AMM where both $x$ and
$y$ can go negative, the value of the LP position also loops on itself and, in the event of borrowing such an LP position during a negative risk free rate,
can exhibit double convexity [\hyperlink{ref-15}{15}].

\subsection{Applications}
The main application happens to be commodities that exhibit holding costs or create deadweight loss if not consumed in time such as electricity, natural gas, oil, and agricultural commodities. For example, excessive sunlight during the day or excessive wind during the night can push electricity prices into the negative domain [\hyperlink{ref-16}{16}] and may be purchased by blockchain miners and data centers positioned near such exchanges managing input costs.

Additionally, the purpose does not necessarily have to apply to the underlying
price going negative. It could apply to an offset representation of the price. For example, an asset can
be worth \$100 and yet its offset price token can be \$0 and now a negative offset price of \$ -1 would mean a
reduction in the price of the underlying to \$99, creating a variety of intriguing non-linear payoffs [Figure4].

Another application of negative prices is to a token that represents the reputation of an individual or an association
(DAO, corporation, non-profit, lobbying group). A negative price would signal a negative reputation.

Consider a scenario where a reputation token does go beyond the zero bound after a large sale, maybe an insider was simply interested in making a quick profit or an anonymous team was aiming to simply deceive uninformed people. By having a zero bound the insider would be able to extract value due to information asymmetry versus the uninformed token purchasers since the price can only approach zero. If a reputation token LP pair creator does not allow for the entry into the negative domain, then it would look incredibly suspicious in a world where reputations can go negative. This can help deal with a lot of the anonymous deceptions one sees where LP pair tokens with a zero bound are created in Uniswap [\hyperlink{ref-17}{17}]. Any swapper would instantly ask why such a token has elected to include a zero bound. If a company/person/anonymous team/DAO truly is committed to their goal, then it should not worry if one’s token price can go negative. The vulnerability of the possibility of a negative price acts as a kind of skin in the game for the anonymous token creator.
\indent

\section{Appendix}

\subsection{A - Negative Liquidity}

\noindent Taking the Uniswap invariant [\hyperlink{ref-18}{18}] with liquidity $L$
\begin{equation}
    x*y=L^2
\end{equation}
introducing price $p$ as $p=y/x$ and $y=px$
\begin{equation}
    x*p*x=L^2
\end{equation}
solving for $x$
\begin{equation}
    x=\sqrt{\frac{L^2}{p}},
\end{equation}
note that a sign appears in front of liquidity
\begin{equation}
    x=\pm\frac{L}{\sqrt{p}}.
\end{equation}

\subsection{B - Liquidity Fingerprint Derivation}

\noindent By taking the second derivative with respect to the square root of price we can derive the liquidity fingerprint in price space. Start with the full invariant
\begin{equation}
\left(x-a\right)^{2}+\left(y-b\right)^{2}=k^{2}.
\end{equation}
Rewrite it as a function of the numeraire
\begin{equation}
    y_{n}=\pm\sqrt{k^{2}-x^{2}+2ax-a^{2}}+b.
\end{equation}
Taking the derivative and negating it
\begin{equation}
    y_{x}=-\frac{dy_{}}{dx}=\pm\frac{\ -x+a}{\sqrt{k^{2}-x^{2}+2ax-a^{2}}}.
\end{equation}
We invert to solve for the price of $x$ due to the symmetric nature of the invariant
 \begin{equation}
    y_{y}=\pm\frac{\sqrt{k^{2}-x^{2}+2ax-a^{2}}}{-x+a}.
\end{equation}
Rearranging the equation for $x$
\begin{equation}
    x=\frac{ay_{y}^{2}+a \pm k\sqrt{y_{y}^{2}+1}}{y_{y}^{2}+1}.
\end{equation}
Rewrite it as a function of price ($\sqrt{p}$) by raising $x^{2}$ and express it as $l(x)$
\begin{equation}
    l(x)=\frac{ax^{4}+a \pm k\sqrt{x^{4}+1}}{x^{4}+1}.
\end{equation}
Taking its derivative to get the liquidity distribution in price space. The distribution happens to have a tail with a Pareto tail index of $\alpha=3$
\begin{equation}
    L(x)=\frac{dl(x)}{dx}=\pm \frac{2kx^{3}}{\left(1+x^{4}\right)^{\frac{3}{2}}}.
\end{equation}
Converting it to tick space $t$ where $t=e^{\frac{x}{2}}$ to get a liquidity fingerprint with heavier tails than the covered call Gaussian distribution:
\begin{equation}
    L(t)=\pm \frac{2ke^{\frac{3t}{2}}}{\left(1+e^{2t}\right)^{\frac{3}{2}}}.
\end{equation}

The value function [\hyperlink{ref-19}{19}] of an LP payoff is given by the Legendre transform with the Greeks for Delta=$\Delta_{c}$, Gamma=$\Gamma_{c}$, and Theta=$\Theta_{c}$ being available in desmos at \url{https://www.desmos.com/calculator/mbohmvmytg}. Expected theta is derived from the relationship of concavity to yield as originally outlined by Louis Bachelier in 1900 [\hyperlink{ref-20}{20}]

\begin{equation}
\Theta_{c}\left(x\right)=E\langle-\frac{\sigma_{iv}^{2}}{2}\Gamma_{c}\left(x\right)\rangle.
\end{equation}
Here implied volatility $\sigma_{iv}$ can be found by looking at the implied volatility at the current ATM strike price or from the relationship between volatility and liquidity at the current tick in Uniswap v3 [\hyperlink{ref-21}{21}].

\subsection{C - Swap functions }
By solving for y in equation (1) for the CCMM we get:
\begin{equation}
y=-\sqrt{k^{2}-x^{2}+2ax-a^{2}}+b
\end{equation}
Adding $\Delta_{x}$ and $\Delta_{y}$ into our swap function to get the quantity of $y$ for quantity $x$:
\begin{equation}
\Delta_{y}=-\sqrt{k^{2}-(x-\Delta_{x})^{2}+2a(x-\Delta_{x})-a^{2}}+b-y
\end{equation}
The swap function, given the symmetric nature of the circle, is the same for $\Delta_{x}$. Alternatively, one can use cosine and sine to rotate along the CCMM and solve for the swap function.

\noindent
For the CSEMM the swap function becomes:
\begin{equation}
y=-\beta\left(\left(1-\left|\frac{x}{\alpha}-1\right|^{\frac{\ln\left(2\right)}{\ln\left(\frac{\alpha}{\alpha-1}\right)}}\right)^{\frac{\ln\left(\frac{\beta}{\beta-1}\right)}{\ln\left(2\right)}}-1\right)
\end{equation}
Adding $\Delta_{x}$ and $\Delta_{y}$ into our swap function
\begin{equation}
\Delta_{y}=-\beta\left(\left(1-\left|\frac{x+\Delta_{x}}{\alpha}-1\right|^{u(\alpha)}\right)^{\frac{1}{u(\beta)}}-1\right)-y
\end{equation}
For $\Delta_{x}$ due to the asymmetric nature of the superellipse our swap function order for $u(x)$ is reversed:
\begin{equation}
\Delta_{x}=-\beta\left(\left(1-\left|\frac{y+\Delta_{y}}{\alpha}-1\right|^{u(\beta)}\right)^{\frac{1}{u(\alpha)}}-1\right)-x
\end{equation}

\subsection{D - Asymmetric Negative Liquidity }
A symmetric invariant in the positive price domain and an asymmetric one in the negative domain can be constructed with a parabola
\begin{equation}
    y=(1-\sqrt{x})^{\beta}.
\end{equation}
where parameter being $\beta>2$ and even. Its liquidity derivation becomes
\begin{equation}
    y_{x}=-\frac{dy_{}}{dx}=\frac{1-\sqrt{x}}{\sqrt{x}}.
\end{equation}
We invert to solve for the price of $x$ due to the symmetric nature of the invariant in the positive price domain
 \begin{equation}
    y_{y}=\frac{\sqrt{x}}{1-\sqrt{x}}.
\end{equation}
Rearranging the equation for $x$
\begin{equation}
    x=\frac{\sqrt{y_{y}}}{1-\sqrt{y_{y}}}.
\end{equation}
Solving for $y_{p}$ we get
 \begin{equation}
    y_{p}=\frac{x^2}{x^2+2x+1}.
\end{equation}
While the invariant is symmetric in the positive price domain, it is not in the negative. If we were to express it as a function of price by squaring the function above, we would lose the asymmetry.

First we solve it as a function of positive price ($\sqrt{p}$) by raising $x^{2}$, expressing it as $l_{+}(x)$
\begin{equation}
    l_{+}(x)=\frac{x^4}{x^{4}+2x^{2}+1}.
\end{equation}
Taking its derivative to get the liquidity distribution $L_{+}$ in the positive domain
\begin{equation}
    L_{+}(x)=\frac{dl(x)}{dx}= \frac{4x^{3}}{\left(1+x^{2}\right)^{3}} \{ x>0\}.
\end{equation}
Converting it to tick space $t$ where $t=e^{\frac{x}{2}}$ we get:
\begin{equation}
    L_{+}(t)= \frac{4e^{\frac{3t}{2}}}{\left(1+e^{t}\right)^{3}} \{ t>0\}.
\end{equation}
To solve it as a function of negative price we can rotate $y_{p}$ in (26) to get
 \begin{equation}
    y_{-p}=\frac{x^2}{x^2-2x+1}.
\end{equation}
Squaring the negative price ($\sqrt{-p}$) by raising $x^{2}$, rewriting it as $l_{-}(x)$
\begin{equation}
    l_{-}(x)=\frac{x^4}{x^{4}-2x^{2}+1}.
\end{equation}
Taking its derivative to get the liquidity distribution $L_{-}$ in the negative domain
\begin{equation}
    L_{-}(x)=\frac{dl(x)}{dx}= -\frac{-4x^{3}}{\left(1-x^{2}\right)^{3}} \{ x<0\}.
\end{equation}
In tickspace we get:
\begin{equation}
    L_{-}(t)= -\frac{-4e^{\frac{3t}{2}}}{\left(e^{t}-1\right)^{3}} \{ t<0\}.
\end{equation}
Connecting the positive and negative liquidity distributions along a unit circle can be seen in Figure 5D.
\newline
\section{Disclaimer}
This paper is for general information purposes only. It does not constitute investment advice or a recommendation
or solicitation to buy or sell any investment and should not be used in the evaluation of the merits of making any
investment decision. It should not be relied upon for accounting, legal or tax advice or investment recommendations.
This paper reflects current opinions of the author and is not made on behalf of any individuals associated with them.
The opinions reflected herein are subject to change without being updated. 

\begin{figure*}[ht!]
\centerline{\includegraphics[width=1\textwidth, keepaspectratio]{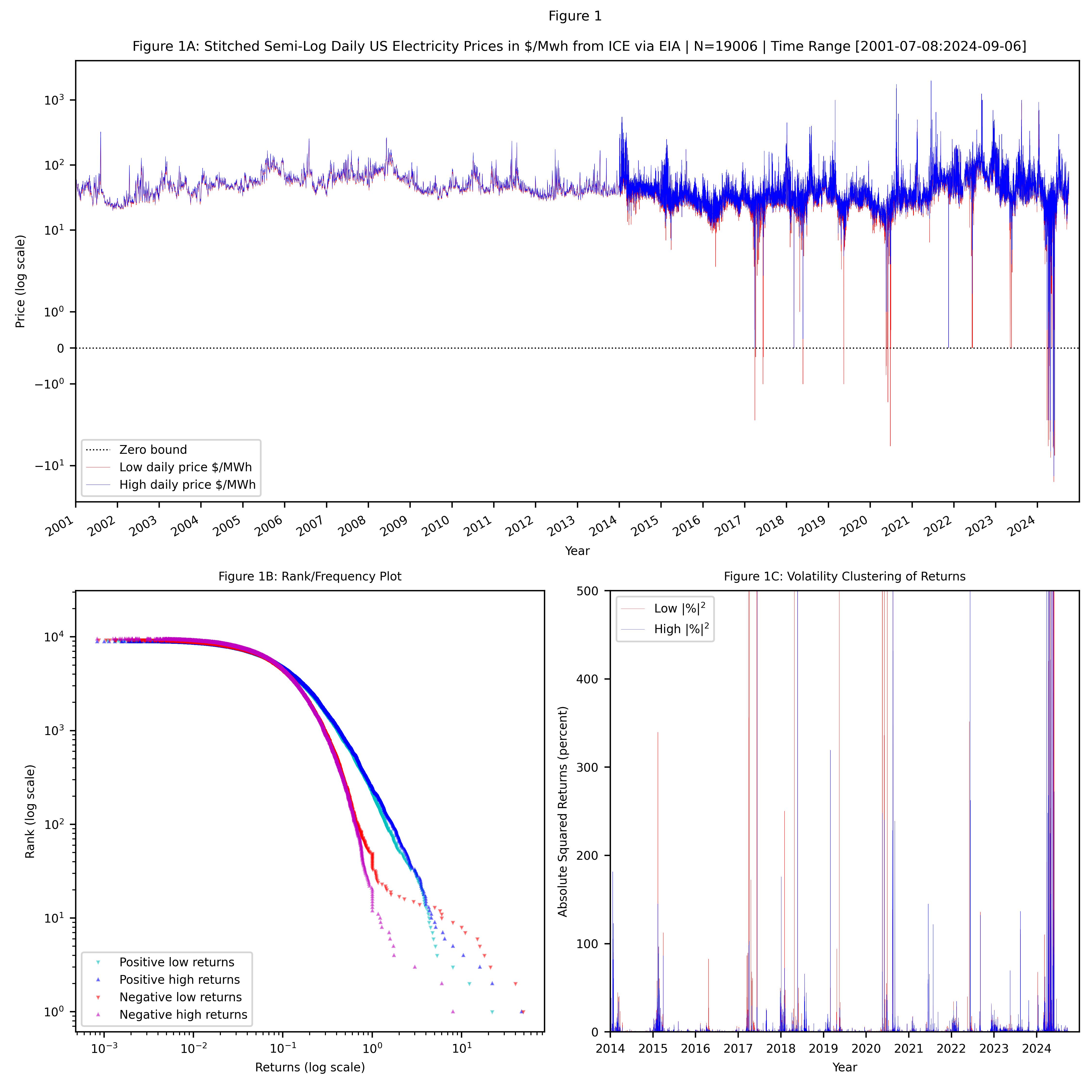}}
\caption{Price dynamics of electricity in the United States. 1A: Daily prices over the past two decades stitched together from the Energy Information Administration [3] containing increasing occurrences of negative prices over time. 1B: Heavy tail asymmetric nature of positive and negative daily returns. 1C: By squaring daily returns we notice volatility clustering over the past decade primarily occurring at the first half of the year followed by a decline with a rise approaching the winter holidays.}
\label{fig1}
\end{figure*}

\begin{figure*}[ht!]
\centerline{\includegraphics[width=0.73\textwidth, keepaspectratio]{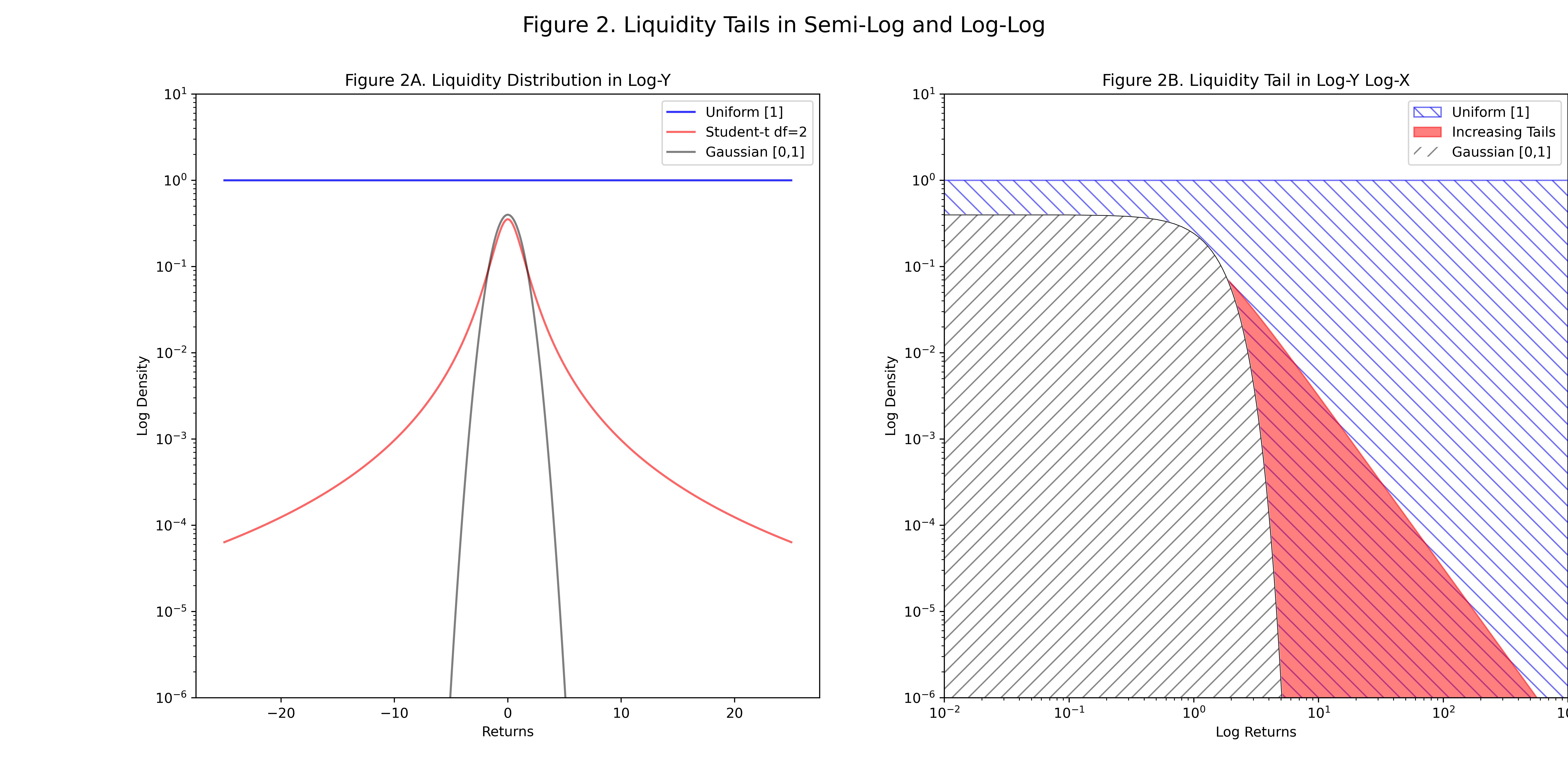}}
\caption{ A CPMM follows a uniform distribution (blue). The liquidity distribution of a covered call RMM [18] follows a Gaussian distribution (black), yet financial assets have the tendency to exhibit fat tail behavior which can cause an LP to under-allocate liquidity in the tails (red area in 2B).}
\label{fig2}
\end{figure*}

\begin{figure*}[ht!]
\centerline{\includegraphics[width=0.73\textwidth, keepaspectratio]{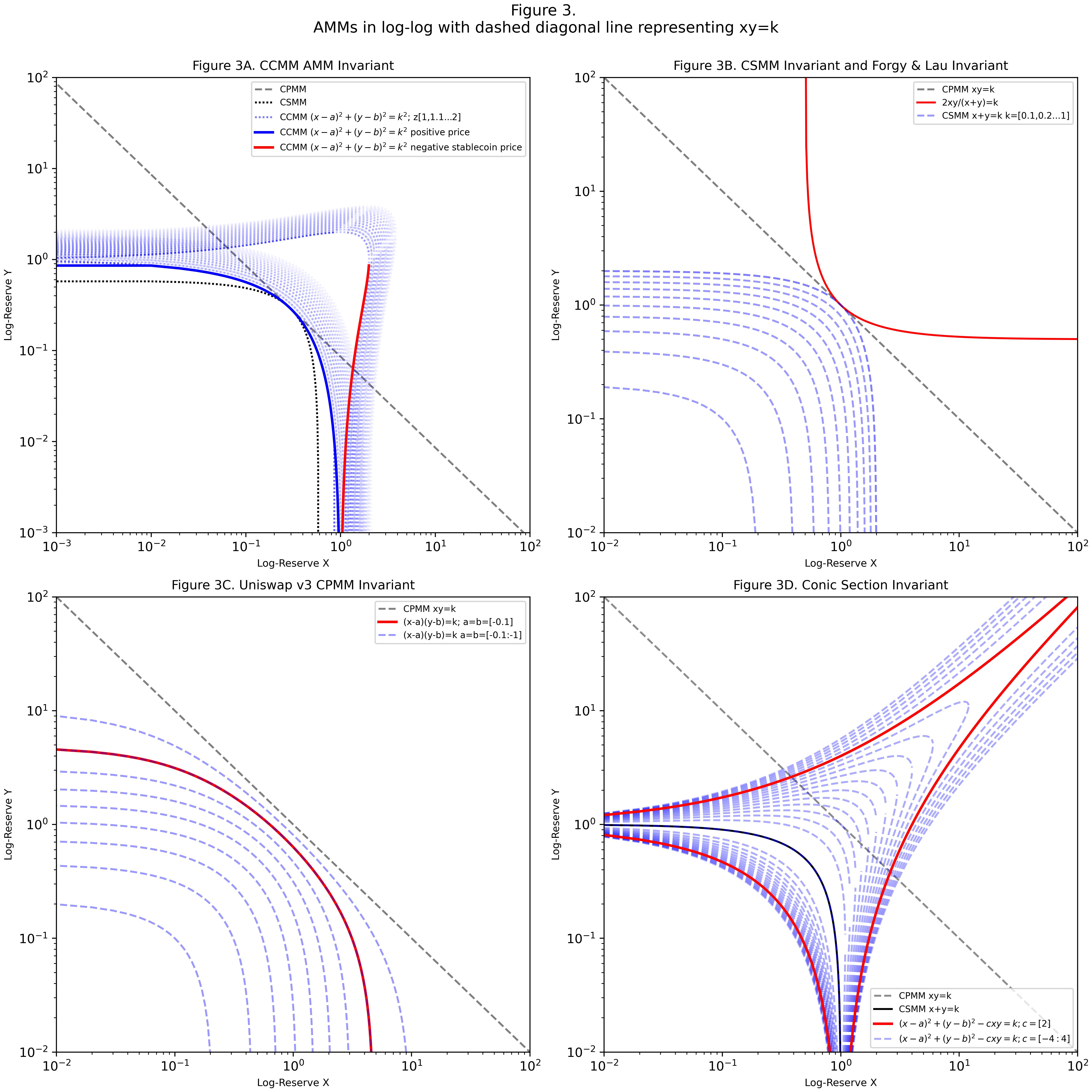}}
\caption{3A: CCMM invariant with red line representing a negative price for an asset priced in a currency with a zero bound
such as a stablecoin. 3B: AMM invariant curve for no impermanent loss as outlined by Eric Forgy and Leo Lau [22] with a CSMM for comparison. 3C: Concentrated liquidity in Uniswap v3. 3D:
Concentration of liquidity in a negatively priced AMM resembles a Uniswap v3 invariant as long as price does not enter negative
territory.}
\label{fig3}
\end{figure*}

\begin{figure*}[ht!]
\centerline{\includegraphics[width=0.73\textwidth, keepaspectratio]{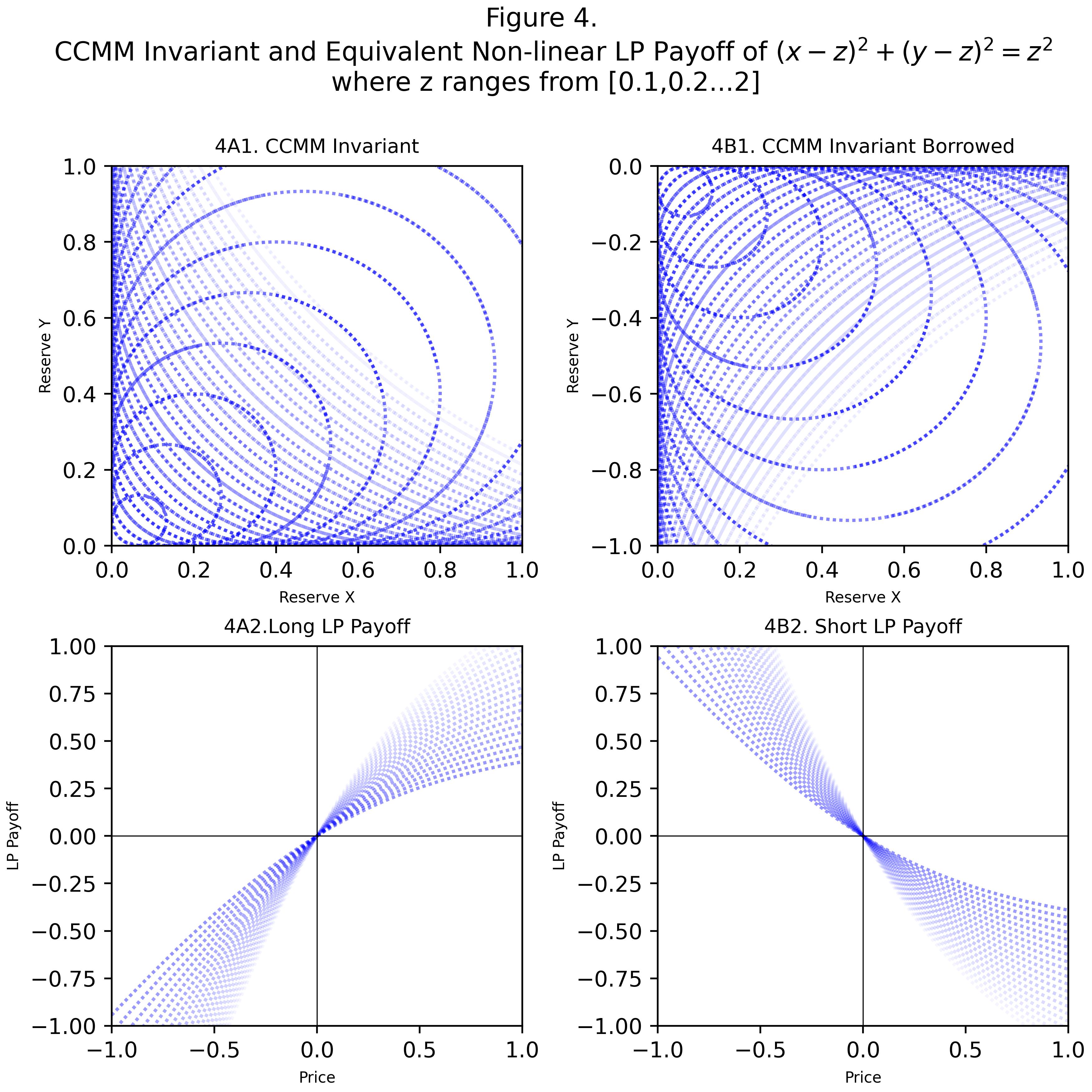}}
\caption{The LP Payoff of a CCMM exhibits concavity when an invariant is constructed. It will exhibit convexity when the LP position is borrowed. The non-linear payoff can further be sharpened and skewed with a CSEMM.
}
\label{fig4}
\end{figure*}

\begin{figure*}[ht!]
\centerline{\includegraphics[width=1\textwidth, keepaspectratio]{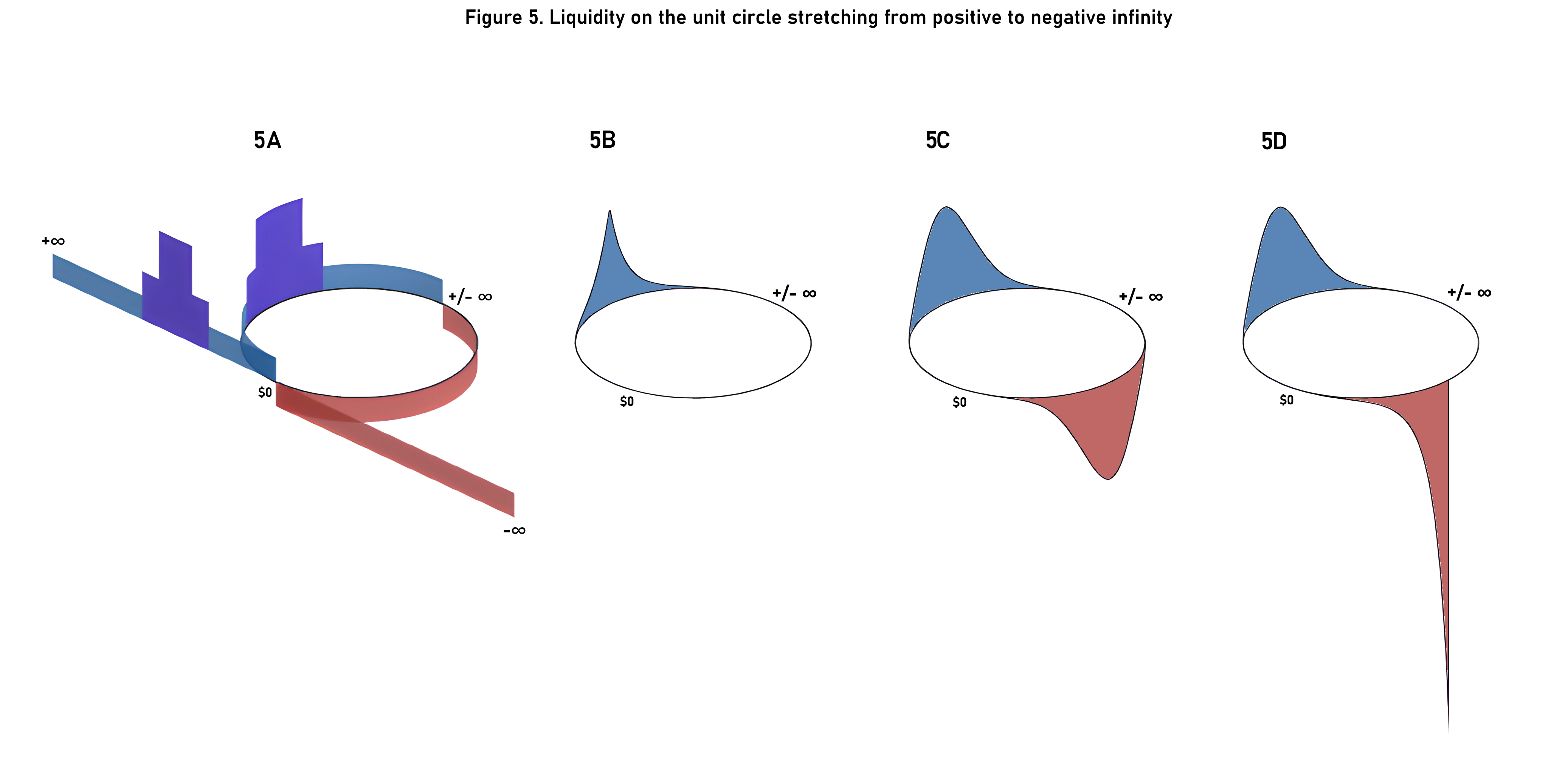}}
\caption{Unit circle with price stretching along it from negative to positive infinity. 5A: Uniswap v2 uniform liquidity in blue stretching across the positive domain and the negative domain in red with two Uniswap v3 concentrated liquidity positions in purple on the positive price domain. 5B: RMM covered call with no negative liquidity given that price cannot go beyond zero. 5C: liquidity distribution of a CCMM with negative liquidity. 5D: liquidity distribution of a parabolic invariant with negative liquidity increasing as price enters the negative domain.}
\label{fig5}
\end{figure*}

\section{References}
\tiny
\noindent [\hypertarget{ref-1}{1}] Madeline Pace. How the COVID-19 Pandemic Plunged Global Oil Prices (2020).\url{https://global.unc.edu/news-story/how-the-covid-19-pandemic-plunged-global-oil-prices/}
\newline
[\hypertarget{ref-2}{2}] Lambert, Emily (2010), The Futures: The Rise of the Speculator and the Origins of the World's Biggest Markets, New York: Basic Books, pp. 240, ISBN 978-0-465-01843-7
\newline
[\hypertarget{ref-3}{3}] U.S. Energy Information Administration. Historical wholesale market data (2024).\url{https://www.eia.gov/electricity/wholesale/#history}
\newline
[\hypertarget{ref-4}{4}] Bloomberg. Trader Error Causes Huge Plunge in Finnish Power Prices (2023). \url{https://www.bloomberg.com/news/articles/2023-11-23/trader-error-causes-huge-plunge-in-finnish-power-prices}
\newline
[\hypertarget{ref-5}{5}] Francis A. Longstaff. Are Negative Option Prices Possible? The Callable U.S. Treasury-Bond Puzzle. Vol. 65, No. 4 (Oct., 1992), pp. 571-592 (22 pages) \url{https://www.jstor.org/stable/2353198}
[\hypertarget{ref-6}{6}]  7 U.S.C. § 13-1 (Pub. L. 85–839, §1, Aug. 28, 1958, 72 Stat. 1013; Pub. L. 111–203, title VII, §721(e)(10), July 21, 2010, 124 Stat. 1672.) \url{https://www.law.cornell.edu/uscode/text/7/13-1}
\newline
[\hypertarget{ref-7}{7}] Hayden Adams, Noah Zinsmeister, and Dan Robinson. Uniswap v2 Core. (2020) \url{https://uniswap.org/whitepaper.pdf}
\newline
[\hypertarget{ref-8}{8}] Hayden Adams et al. Uniswap v3. (2021) \url{https://uniswap.org/whitepaper-v3.pdf}
\newline
[\hypertarget{ref-9}{9}] Alexander Angel, et al. Financial Virtual Machine. (2022) \url{https://www.primitive.xyz/papers/yellow.pdf}
\newline
[\hypertarget{ref-10}{10}] Gabaix, X., Gopikrishnan, P., Plerou, V. et al. A theory of power-law distributions in financial market fluctuations. Nature 423, 267–270 (2003). \url{https://doi.org/10.1038/nature01624}
\newline
[\hypertarget{ref-11}{11}] Jean-Philippe Bouchaud. Power-laws in Economy and Finance: Some Ideas from Physics (2018).\url{https://doi.org/10.48550/arXiv.cond-mat/0008103}
\newline
[\hypertarget{ref-12}{12}]  Naureen S Malik. Negative Power Prices? Blame the US Grid for Stranding Renewable Energy. Retrieved from Bloomberg November 20, 2023 \url{https://www.bloomberg.com/news/articles/2022-08-30/trapped-renewable-energy-sends-us-power-prices-below-zero}
\newline
[\hypertarget{ref-13}{13}]  Examen des différentes méthodes employées pour résoudre les problèmes de géométrie. Gabriel Lam\'e (1818) \url{https://gallica.bnf.fr/ark:/12148/bpt6k92728m/f119.item.texteImage}
[\hypertarget{ref-14}{14}]  CME Group. Switch to Bachelier Options Pricing Model - Effective April 22, 2020.  (2020) \url{https://www.cmegroup.com/notices/clearing/2020/04/Chadv20-171.html}
\newline
[\hypertarget{ref-15}{15}] Artemis Capital Management. Volatility at World's End: Deflation, Hyperinflation, and the Alchemy of Risk (2012). \url{https://www.asx.com.au/content/dam/asx/investors/investment-options/vix/volatility-at-worlds-end.pdf}
\newline
[\hypertarget{ref-16}{16}] Brian Bartholomew. Twitter (Feb. 14, 2024). \url{https://twitter.com/BPBartholomew/status/1757991908578578570/photo/1}
\newline
[\hypertarget{ref-17}{17}] Bruno Mazorra, Victor Adan, Vanesa Daza. Do not rug on me: Zero-dimensional Scam Detection (2022). \url{https://doi.org/10.48550/arXiv.2201.07220}
\newline
[\hypertarget{ref-18}{18}] Dan Robinson. 2021. Uniswap v3: The Universal AMM. \url{https://www.paradigm.xyz/2021/06/uniswap-v3-the-universal-amm}
\newline
[\hypertarget{ref-19}{19}] Guillermo Angeris, Alex Evans, and Tarun Chitra. Replicating market makers (2021). \url{https://doi.org/10.48550/arXiv.2111.13740}
\newline
[\hypertarget{ref-20}{20}] Louis Bachelier. Theorie de la Speculation (1900), page 100. \url{https://archive.org/details/bachelier-theorie-de-la-speculation/page/n5/mode/2up}
\newline
[\hypertarget{ref-21}{21}] Guillaume Lambert. Yewbow. Retrieved Nov 20, 2023 from \url{https://info.yewbow.org/#/pools}
\newline
[\hypertarget{ref-22}{22}] Eric Forgy and Leo Lau. A family of multi-asset automated market makers.  Retrieved Oct 2, 2024 from \url{https://doi.org/10.48550/arXiv.2111.08115}
\newline
{\ }
{\ }

\end{document}